# Quantum Simulation of Molecular Dynamics Processes − A Benchmark Study Using Classical Simulator and Present-Day Quantum Hardware

Tamila Kuanysheva,[1] Brian Kendrick,[2] Lukasz Cincio,[3] and Dmitri Babikov[1*]

**Affiliations:**

[1]*Chemistry Department, Marquette University, Milwaukee, Wisconsin 53201-1881, USA*
[2]*Los Alamos National Laboratory, Theory Division, T-1, Los Alamos, New Mexico 87545, USA*
[3]*Los Alamos National Laboratory, Theory Division, T-4, Los Alamos, New Mexico 87545, USA*
\* E-mail: dmitri.babikov@mu.edu

**Abstract:** We explore how the fundamental problems in quantum molecular dynamics can be modelled using classical simulators (emulators) of quantum computers and the actual quantum hardware available to us today. The list of problems we tackle includes propagation of a free wave packet, vibration of a harmonic oscillator, and tunneling through a barrier. Each of these problems starts with the initial wave packet setup. Although Qiskit provides a general method for initializing wavefunctions, in most cases it generates deep quantum circuits. While these circuits perform well on noiseless simulators, they suffer from excessive noise on quantum hardware. To overcome this issue, we designed a shallower quantum circuit for preparing a Gaussian-like initial wave packet, which improves the performance on real hardware. Next, quantum circuits are implemented to apply the kinetic and potential energy operators for the evolution of a wavefunction over time. The results of our modelling on classical emulators of quantum hardware agree perfectly with the results obtained using the traditional (classical) methods. This serves as a benchmark and demonstrates that the quantum algorithms and Qiskit codes we developed are accurate. However, the results obtained on the actual quantum hardware available today, such as IBM's superconducting qubits and IonQ's trapped ions, indicate large discrepancies due to hardware limitations. This work highlights both the potential and challenges of using quantum computers to solve fundamental quantum molecular dynamics problems.

# I. INTRODUCTION

Quantum computing carries great potential for an efficient prediction of physical and chemical properties of molecules.[1–5] This is the case simply because many molecular properties are determined by quantum mechanics and many molecular processes are governed by quantum mechanical principles. Relevant examples include molecular energy transfer,[6] molecule-light interaction,[7–9] electron addition to (or removal from) a molecule,[10] formation and breakage of chemical bonds,[11–17] to name just a few.

While a very significant effort has already been devoted to the quantum computation of molecular electronic structure,[18–29] considerably less attention was paid to the simulations of quantum molecular dynamics. Quantum molecular dynamics describes the motion of atoms during molecular transformations and encompasses such phenomena as molecular vibrations and rotations, bond breaking in the course of a chemical reaction, collisions of two molecules, *etc.* – all driven by the time-dependent Schrodinger equation. One of the earliest proposals to employ quantum computing for the description of inelastic collisions was based on a mixed quantum-classical approach and utilized so-called analog quantum computing.[6,30] This methodology was later re-formulated using a general quantum computing approach that utilizes a universal set of quantum logic gates (sometimes called digital quantum computing).[31,32] A mixed quantum-classical approach was further explored in the variational time propagation algorithms.[33] An alternative algorithm for the description of inelastic molecular scattering within the framework of the time-independent Schrodinger equation, and using quantum computing to solve it, has also been proposed.[34] A practical calculation of bound vibrational states of molecules below the dissociation threshold, and scattering resonances above it, was demonstrated using the D-Wave quantum annealer.[35,36] A quantum annealer algorithm was also proposed and implemented to propagate trajectories for large-amplitude molecular vibrations and for bond breaking.[37] Finally, grid methods for the encoding of time-dependent wavefunctions on quantum computers were developed by several authors to describe molecular vibrations,[38,39] electron scattering,[10] and quantum tunneling.[40–42]

In this paper we review all elements of the quantum computing approach needed to solve the time-dependent Schrodinger equation using a grid-representation of the wavefunction. We apply this methodology to solve three fundamental problems encountered in the numerical simulations of quantum molecular dynamics, which includes: 1) propagation of a wave packet on a flat landscape of potential energy, typical to the entrance channel of a chemical process that describes reagents; 2) quantum tunneling through a barrier (as shown in the graphical abstract) that describes the activated transition state of a chemical process, and finally 3) oscillations of the molecular bond length that describes the vibrationally excited product of a reaction. For these three problems we report the results obtained by running quantum algorithms on a classical emulator of quantum hardware (Qiskit SDK primitives)[43] and on the actual present day quantum hardware (such as IBM Brisbane, IBM Torino, and IonQ Aria 1).[44] To the best of our knowledge, this is the first systematic comparison of several quantum processors currently available to the community. Examples of quantum codes written in Qiskit [43,45] are available to readers in the supplementary material. These codes can be used as a good starting point by those who would like to learn how to write their first quantum code and how to run the quantum dynamics calculations on a simulator of a quantum computer, or on the actual quantum processor.

# II. COMPUTATIONAL METHODOLOGY

## II-A. Wavefunction Encoding

The time-dependent wavefunction of the system $\psi(r,t)$, sometimes called a wave packet, is a function of time $t$ but it also depends on a translational coordinate $r$ that can represent, for example, the length of a molecular bond, or the position along a generalized reaction path. The potential energy $V$ of the system is also a function of this variable, $V(r)$. This degree of freedom is discretized by introducing a grid of points $r_m$ along $r$. Here we consider a simple equidistant grid with a constant step size $\Delta r = (r_{max} - r_{min})/M$ where $M$ is the number of grid points, and $r_m = r_{min} + (1/2 + m)\Delta r$ with $m$ in the range $0 \leq m \leq M-1$. A variable-step grid $\Delta r(r)$, optimized for a particular shape of the molecular potential $V(r)$, can also be implemented.[46–48]

The wavefunction is represented by a set of its values at these grid points, $\psi_m(t) = \psi(r_m, t)$, which is often called a discrete variable representation (DVR),[49] and is known to be mathematically equivalent to the expansion of a wave function over a finite basis set (FBR) of sinc functions.[10,28] Similarly, for the potential energy function we have $V_m = V(r_m)$, as shown in Fig. 1.

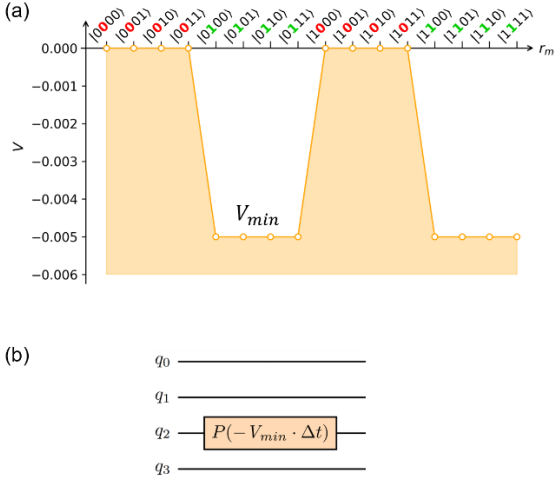

**Figure 1**: a) Mapping of the coordinate grid onto the states of a quantum computer with four-qubits, and a double-well potential defined on this grid. The states labelled with red and green digits correspond to $V = 0$ and $V = V_{min}$, respectively, where $V_{min}$ is well depth. b) Quantum circuit to implement the potential energy operator for this model (explained in section II-C).

On a quantum computer with $n$ qubits, the $2^n = M$ states of the quantum register are available and can be encoded with the values of wavefunction $\psi_m$ at the $M$ points of the grid. Before diving into further details, one should mention that Qiskit follows a little-endian convention for qubit labeling, where qubit 0 corresponds to the rightmost digit in the *ket* notation (least significant bit, LSB), while qubit $n-1$ corresponds to the leftmost digit in the *ket* notation (most significant bit, MSB).[50] For example, a quantum computer with four qubits, $n = 4$, has $M = 16$ states that can be used to represent a grid of 16 points, with the corresponding mapping of the wavefunction $\psi_m$, where $0 \leq m \leq 15$. These qubit states are listed along the $r$-axis in Fig. 1, starting with the $|0000\rangle$ state that corresponds to the first point of the grid $r_0$, $|0001\rangle$ to the second point $r_1$, *etc.*, up to the state $|1111\rangle$ that corresponds to the last point of the grid $r_{M-1}$. To label these states in a more concise way we will use the $|m\rangle$ notation with a *ket* where $m$ labels the states of the multi-qubit system, namely, $|0\rangle$ corresponds to $|0000\rangle$, $|1\rangle$ corresponds to $|0001\rangle$, and so on, up to $|15\rangle$, which corresponds to $|1111\rangle$. The basis of states for a register of 4 qubits can also be written as: $|m\rangle = |q_3\rangle \otimes |q_2\rangle \otimes |q_1\rangle \otimes |q_0\rangle$, where $|q_j\rangle$ represents the basis of two states 0 and 1 of the qubit number $j$, and the symbol $\otimes$ denotes a tensor product. For an arbitrary number of qubits $n$, we have:

$$|m\rangle = \bigotimes_{j=n-1}^{0} |q_j\rangle \quad (1)$$

It is worth mentioning that the ket labels 0000, 0001, 0010, 0011, …, 1111 in Fig. 1(a) represent binary numbers corresponding to the integer numbers 0, 1, 2, 3, …, 15.

### II-B. Split Operator Propagation Method

Throughout this paper, we adopt atomic units, setting $\hbar = 1$ for simplicity. Formally, the evolution of wave function is obtained by applying the time propagation operator to the initial wave function:

$$\psi(r, t) = e^{-i\widehat{H}t} \psi(r, 0) \quad (2)$$

where $\widehat{H} = \widehat{T} + \widehat{V}$ is the Hamiltonian operator, $\widehat{V} = V(r)$, $\widehat{T} = \widehat{p}^2/2\mu$ is kinetic energy operator and $\mu$ is the reduced mass. One of the simplest practical ways to do this is by using the Split-Operator method,[51–53] which involves, first, approximating the propagator over the global time interval $[0, t_{\text{fin}}]$ as a product of propagators over shorter time intervals $\Delta t$:

$$e^{-i\widehat{H}t_{\text{fin}}} \approx (e^{-i\widehat{H}\Delta t})^N \quad (3)$$

where $N$ is the number of time-steps, $N\Delta t = t_{\text{fin}}$. Next, each short-time propagator is approximated as a symmetrized product of kinetic and potential energy components:[52,54]

$$e^{-i\widehat{H}\Delta t} \approx e^{-i\frac{V}{2}\Delta t} e^{-i\frac{\widehat{p}^2}{2m}\Delta t} e^{-i\frac{V}{2}\Delta t} + O(\Delta t^3). \quad (4)$$

Figure 2 gives a schematic representation of the Split-Operator method within a quantum computing framework.[40–42,54–58] A detailed explanation of each component is provided in subsequent sections.

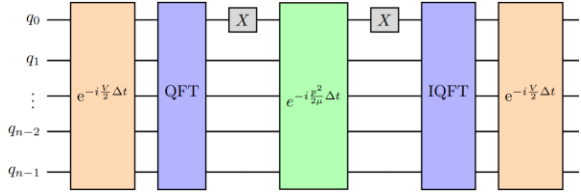

**Figure 2**: Quantum circuit for one time-step of a split-operator propagation using the quantum Fourier transform without the swap gates. The number of qubits is $n$.

## II-C. Implementing Potential Energy Operator: Double-Well Potential

In the coordinate representation described above the potential energy operator $V(r)$ is diagonal, so, the application of the potential energy term of the time evolution operator is relatively straightforward, $\psi(r, \Delta t) = e^{-iV(r)\Delta t}\psi(r, 0)$. This means that if the wavefunction is represented by a grid of points $r_m$, then the wavefunction $\psi_m$ at each point $m$ evolves independently from those at other points of the grid by acquiring an appropriate phase shift $\varphi_m$:

$$\psi_m(\Delta t) = e^{i\varphi_m}\psi_m(0) \qquad (5)$$

where $\varphi_m = -V_m \Delta t$, with $V_m = V(r_m)$ as introduced above. The amplitude of wavefunction $|\psi_m|$ remains unaffected by this operation.

Since the potential energy only affects the phase of the wavefunction (not its amplitude), it can be implemented efficiently on a quantum computer using the phase gate $P(\varphi)$ that introduces a phase shift $\varphi$ between the two states of a qubit.[45,59,60] In the matrix form:

$$P(\varphi) = \begin{vmatrix} 1 & 0 \\ 0 & e^{i\varphi} \end{vmatrix}$$

Now consider the double-well potential depicted in Fig. 1. Note that in this case the phase shift $\varphi = -V_{min} \Delta t$ should only be applied to those states of the quantum register where the second most significant qubit is in state 1 (indicated by green color in Fig. 1). For the remaining states, where the second most significant qubit is in state 0 (indicated by red color in Fig. 1), no phase shift is needed because these states map onto the points of the grid where the potential is zero ($V = 0$ at the top of the barrier between the wells, see Fig. 1).

This can be achieved by applying the $P(\varphi)$ gate to the second most significant qubit ($q_2$ in the four-qubit example of Fig. 1), thereby influencing the relevant subset of computational states:[40,41]

$$e^{-i\hat{V}\Delta t} = I \otimes P(\varphi) \otimes I \otimes I \qquad (6)$$

where $\varphi = -V_{min} \Delta t$ and $I$ is an identity operator given by the $2 \times 2$ identity matrix. The quantum circuit to implement Eq. (6) is given in Fig. 1b. As one can easily check, in the matrix form this transformation is equivalent to the diagonal matrix $16 \times 16$ presented in Fig. 3. This operation effectively partitions the system into four spatial regions seen in Fig. 1a, selectively applying the phase shift to the relevant basis states, thereby modeling a double-well potential landscape.

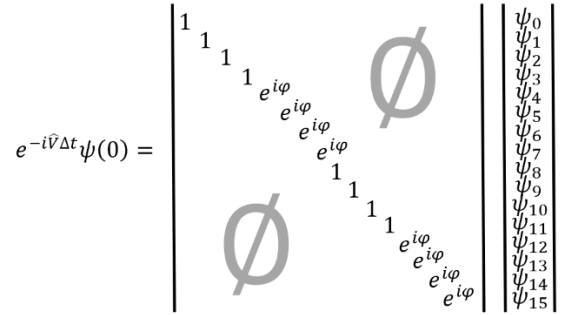

**Figure 3**: Representation of the potential energy operator in the double-well potential problem in terms of the matrix acting on the vector of 4-qubit states. Diagonal elements of the matrix describe phase shifts due to potential in Fig. 1a.

## II-D. Implementing Potential Energy Operator: Harmonic Oscillator

To model the harmonic oscillator problem on a quantum computer we designed a quantum circuit implementing the potential energy operator for a quadratic potential energy function (see Fig. 4):

$$V(r_m) = \tfrac{1}{2}k(r_m - r_{eq})^2 \qquad (7)$$

where $r_{eq}$ is the equilibrium position, $k = \mu\omega^2$ is the force constants, and $\omega$ is the harmonic frequency.

Substitution of $r_m = r_{min} + (1/2 + m)\Delta r$ into Eq. (7), and using Eq. (7) in the potential energy operator, leads to the following result:

$$e^{-i\hat{V}\Delta t} = e^{-iV(r_m)\Delta t} = e^{i(m^2\alpha + m\beta + \gamma)} \quad (8)$$

where we introduced the following rotation angles:

$$\alpha = -k\Delta r^2 \Delta t / 2,$$

$$\beta = \alpha \cdot \left(\frac{2r_{min} - 2r_{eq} + \Delta r}{\Delta r}\right),$$

$$\gamma = \alpha \cdot \left(\frac{2r_{min} - 2r_{eq} + \Delta r}{2\Delta r}\right)^2$$

We see that in the harmonic oscillator model the amount of phase shift $\varphi$, introduced by the potential energy operator in Eq. (8), depends on the location of the point along the $r$-grid: $\varphi_m = m^2\alpha + m\beta + \gamma$. As one could expect, this dependence is quadratic in $m$.

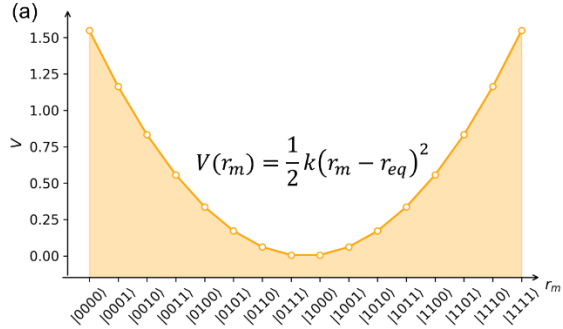

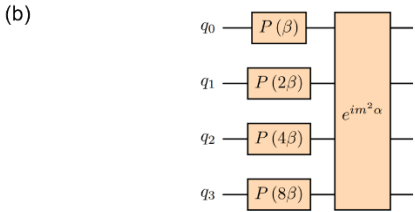

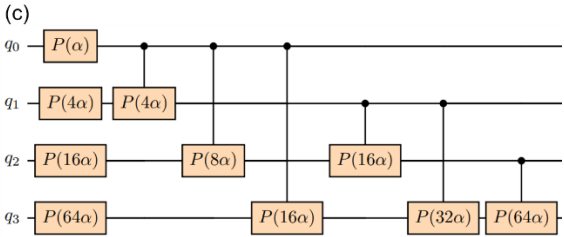

**Figure 4**: a) Mapping of the coordinate grid onto the states of a quantum computer with four-qubits, and a harmonic potential; b) Quantum circuit to implement the potential energy operator for this model; c) Quantum circuit to implement the quadratic term of the potential energy operator in part (b). Black dots indicate control qubits of the two-qubit gates.

Equation (8) can be expressed as a product of three rotations. The last term $e^{i\gamma}$ represents a global phase shift, which is irrelevant. The second phase term in Eq. (8), linear in the index $m$, can be expressed, using its binary representation, through the qubit number $j$ as follows:

$$m = \sum_{j=n-1}^{0} q_j 2^j \quad (9)$$

where $q_j = 0$ if qubit $j$ is in state 0 and $q_j = 1$ if qubit $j$ is in state 1. This allows us to rewrite the second phase term in Eq. (8) as follows:

$$e^{i(m\beta)}|m\rangle = \exp\left\{i\beta \sum_{m=n-1}^{0} q_j 2^j\right\}|m\rangle \quad (10)$$

$$= \bigotimes_{j=n-1}^{0} e^{i(q_j 2^j \beta)}|q_j\rangle$$

It is important to emphasize the utility of Eq. (10) -- it expresses the global operator $e^{i(m\beta)}$ that acts on the entire wavefunction $|m\rangle$ through phase gates that act on the individual qubits $|q_j\rangle$. For the 4-qubit example represented by Fig. 4, this operator is:

$$e^{i(m\beta)} = P(8\beta) \otimes P(4\beta) \otimes P(2\beta) \otimes P(\beta) \quad (11)$$

This expression shows that the linear term in Eq. (8) can be implemented using single-qubit phase shift gates, with rotation angles determined by the respective binary weights of each qubit in the register. The values of rotation angles (multiples of angle $\beta$) are determined by the parameters of the model (the potential and the grid), as introduced above. These four gates represent the initial step of the circuit given in Fig. 4b.

Finally, the leading term in Eq. (8), quadratic in $m$, can be re-expressed using the binary representation of $m^2$:

$$m^2 = \left(\sum_{j=n-1}^{0} q_j 2^j\right)^2$$

$$= \sum_{j=n-1}^{0} q_j 2^{2j} + \sum_{k=n-1}^{1} \sum_{j<k}^{0} 2 q_j q_k 2^{j+k} \qquad (12)$$

which allows us to factorize this operator as:

$$e^{i(m^2\alpha)}|m\rangle$$
$$= \bigotimes_{j=n-1}^{0} e^{i(\alpha q_j 2^{2j})}|q_j\rangle \bigotimes_{\substack{k=n-1,\\j<k}}^{1} e^{i(\alpha 2 q_j q_k 2^{j+k})}|q_k q_j\rangle \qquad (13)$$

In the 4-qubit case, the first part of this operator is applied by phase gates acting on the individual qubits: $P(64\alpha) \otimes P(16\alpha) \otimes P(4\alpha) \otimes P(\alpha)$, as shown in Fig. 4c. The second part of Eq. (13) can be implemented on a quantum computer using two-qubit gates known as controlled phase gates,[43,60] or $CP(\varphi)$. A controlled-phase gate introduces a phase shift $\varphi$ into the target qubit if the control qubit is in the state 1, but leaves the target qubit unchanged if the control qubit is in state 0: $CP(\varphi) = I \otimes |0\rangle\langle 0| + P(\varphi) \otimes |1\rangle\langle 1|$. Its matrix form, in the $4 \times 4$ space of two qubits, is:

$$CP(\varphi) = \begin{vmatrix} 1 & 0 & 0 & 0 \\ 0 & 1 & 0 & 0 \\ 0 & 0 & 1 & 0 \\ 0 & 0 & 0 & e^{i\varphi} \end{vmatrix} \qquad (14)$$

where

$$|0\rangle\langle 0| = \begin{vmatrix} 1 \\ 0 \end{vmatrix} |1 \quad 0| = \begin{vmatrix} 1 & 0 \\ 0 & 0 \end{vmatrix}$$

$$|1\rangle\langle 1| = \begin{vmatrix} 0 \\ 1 \end{vmatrix} |0 \quad 1| = \begin{vmatrix} 0 & 0 \\ 0 & 1 \end{vmatrix}$$

For the 4-qubit example given by Fig. 4a, six $CP(\varphi)$ gates are needed, as shown in the circuit presented in Fig. 4c. The values of rotation angles $\varphi$ (multiples of angle $\alpha$) are determined by the parameters of the model, as introduced above. It is somewhat tedious, but possible to show, that the overall circuit presented in Fig. 4b and 4c for the potential energy operator in the harmonic oscillator model is equivalent to the diagonal matrix presented in Fig. 5. The quantum circuit for a general $n$-qubit encoding of the harmonic potential is given in the supplementary material.

$$e^{-i\hat{V}\Delta t}\psi(0) = \begin{vmatrix} e^{i(0)} & & & & & & & & & & & & & & & \\ & e^{i(\alpha+\beta)} & & & & & & & & & & & & & & \\ & & e^{i(4\alpha+2\beta)} & & & & & & & & & & & & & \\ & & & e^{i(9\alpha+3\beta)} & & & & & & & & & \varnothing & & & \\ & & & & e^{i(16\alpha+4\beta)} & & & & & & & & & & & \\ & & & & & e^{i(25\alpha+5\beta)} & & & & & & & & & & \\ & & & & & & e^{i(36\alpha+6\beta)} & & & & & & & & & \\ & & & & & & & e^{i(49\alpha+7\beta)} & & & & & & & & \\ & & & & & & & & e^{i(64\alpha+8\beta)} & & & & & & & \\ & & & & & & & & & e^{i(81\alpha+9\beta)} & & & & & & \\ & & & & & & & & & & e^{i(100\alpha+10\beta)} & & & & & \\ & & & & & & & & & & & e^{i(121\alpha+11\beta)} & & & & \\ & \varnothing & & & & & & & & & & & e^{i(144\alpha+12\beta)} & & & \\ & & & & & & & & & & & & & e^{i(169\alpha+13\beta)} & & \\ & & & & & & & & & & & & & & e^{i(196\alpha+14\beta)} & \\ & & & & & & & & & & & & & & & e^{i(225\alpha+15\beta)} \end{vmatrix} \begin{vmatrix} \psi_0 \\ \psi_1 \\ \psi_2 \\ \psi_3 \\ \psi_4 \\ \psi_5 \\ \psi_6 \\ \psi_7 \\ \psi_8 \\ \psi_9 \\ \psi_{10} \\ \psi_{11} \\ \psi_{12} \\ \psi_{13} \\ \psi_{14} \\ \psi_{15} \end{vmatrix}$$

**Figure 5**: Representation of the potential energy operator in the harmonic oscillator problem in terms of the matrix acting on the vector of 4-qubit states. Diagonal elements of the matrix describe phase shifts due to the potential in Fig. 4a.

### II-E. Implementing Kinetic Energy Operator

In the position representation (here $r$ coordinate), the kinetic energy operator $\hat{T}$ appears as a non-diagonal matrix, making its implementation more challenging. However, in the momentum space, the operator $\hat{T}$ becomes diagonal and can be implemented similar to the potential energy operator discussed above, by applying the time-propagation operator $e^{-i\hat{T}\Delta t}$ to each point of the grid independently. On classical computers, this process involves transforming the wave function from the position representation $\psi(r_m)$ to the momentum representation, applying the diagonal kinetic energy operator in the momentum representation, and then transforming back to the position representation:[54,57,61]

$$e^{-i\hat{T}\Delta t}\psi(r_m) = \hat{Z}\, e^{-i\frac{p_m^2}{2\mu}\Delta t}\, \hat{Z}^\dagger \psi(r_m) \qquad (15)$$

where $\hat{Z}^\dagger$ represents the transformation between position and momentum representations. Since an equidistant grid corresponds to the Fourier basis, the transformation $\hat{Z}^\dagger$ corresponds to the discrete Fourier transform (DFT), which on classical computers can be efficiently implemented using the fast Fourier transform (FFT) algorithm, and $\hat{Z}$ is the inverse FFT. On a quantum computer, the DFT is naturally implemented using the Quantum Fourier Transform (QFT), which is an essential quantum algorithm available on platforms such as IBM's Qiskit.[43,45] Figure 6 represents the quantum circuits for the QFT and its inverse (IQFT) of four qubits.

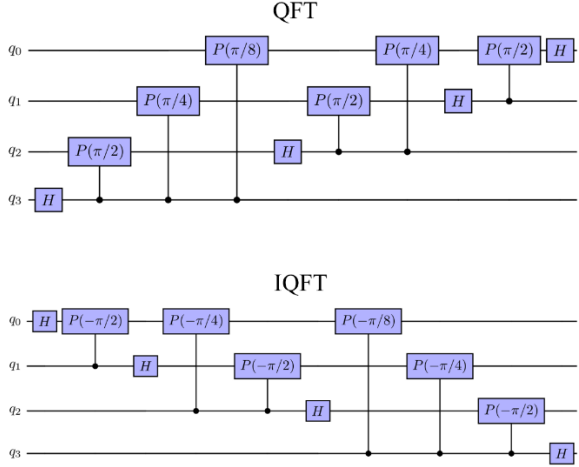

**Figure 6**: Quantum circuit implementation of the direct and inverse QFT in IBM's Qiskit for a 4-qubit system. The version without swap gates is shown, as used in this work.

The discretized momentum values are defined as:

$$p_m = -p_{max} + m\Delta p \quad (16)$$

where $M$ is the number of points and $m$ labels points of the grid as before, $p_{max} = M\Delta p/2$ and $\Delta p$ is the step-size in the momentum representation. From the Nyquist theorem in the theory of discrete Fourier transform,[45,62] we have $\Delta r \Delta p = 2\pi/M$. Then, from Eq. (16) we obtain:

$$p_m = \frac{2\pi}{r_{max}-r_{min}}\left(m - \frac{M}{2}\right) \quad (17)$$

$$p_m^2 = \left(\frac{2\pi}{r_{max}-r_{min}}\right)^2 (m^2 - mM + M^2/4) \quad (18)$$

and finally:

$$e^{-i\, p_m^2/(2\mu)\,\Delta t} = e^{i(m^2\theta + m\phi + \delta)} \quad (19)$$

which is analogous to Eq. (8). Here the following rotation angles were introduced:

$$\theta = -\left(\frac{2\pi}{r_{max}-r_{min}}\right)^2 \frac{\Delta t}{2\mu},$$

$$\phi = -\theta \cdot M,$$

$$\delta = \theta \cdot M^2/4.$$

As was stated earlier, global phase factors like $e^{i\delta}$ are irrelevant since they do not affect measurement probabilities in quantum computations. The other two terms in Eq. (19), linear and quadratic in $m$, can be implemented on a quantum computer in a way analogous to what was described for the harmonic potential in the previous section, namely:

$$e^{i(m\phi)}|m\rangle = \bigotimes_{j=n-1}^{0} e^{i(q_j 2^j \phi)}|q_{n-1-j}\rangle \quad (20)$$

$$e^{i(m^2\theta)}|m\rangle = \bigotimes_{j=n-1}^{0} e^{i(\theta q_j 2^{2j})}|q_{n-1-j}\rangle \quad (21)$$

$$\bigotimes_{\substack{k=n-1,\\ j<k}}^{1} e^{i(\theta 2 q_j q_k 2^{j+k})}|q_{n-1-k} q_{n-1-j}\rangle$$

One can see that these expressions are analogous to Eqs. (10) and (13). Figure 7 illustrates the quantum circuits for these operators in the case of four qubits. The quantum circuit for the kinetic energy operator in a general $n$-qubit case is given in the supplementary material.

Since we apply the QFT without swap gates, the resulting qubit states appear in a shuffled order (as in Eq. (22) below) due to the inherent bit-reversal effect of the transform. To obtain the order consistent with the monotonic discretization of Eq. (16), an $X$-gate should be applied to the first qubit $q_0$ immediately after the QFT and before the IQFT, as illustrated in Fig. 2. This operation effectively implements a pairwise reordering of the basis states $|m\rangle$, ensuring the correct mapping of the computational states. In the case of four qubits:

$$(I \otimes I \otimes I \otimes X)\begin{pmatrix}\psi_1\\ \psi_0\\ \psi_3\\ \psi_2\\ \psi_5\\ \psi_4\\ \psi_7\\ \psi_6\\ \psi_9\\ \psi_8\\ \psi_{11}\\ \psi_{10}\\ \psi_{13}\\ \psi_{12}\\ \psi_{15}\\ \psi_{14}\end{pmatrix} = \begin{pmatrix}\psi_0\\ \psi_1\\ \psi_2\\ \psi_3\\ \psi_4\\ \psi_5\\ \psi_6\\ \psi_7\\ \psi_8\\ \psi_9\\ \psi_{10}\\ \psi_{11}\\ \psi_{12}\\ \psi_{13}\\ \psi_{14}\\ \psi_{15}\end{pmatrix} \quad (22)$$

where the $X$-gate in the matrix form is used:

$$X = \begin{vmatrix} 0 & 1 \\ 1 & 0 \end{vmatrix}$$

This goal can also be achieved by introducing the so called cQFT, as it was proposed in the literature.[38]. Both of these methods are different from the standard QFT with swap gates,[62] where the values of momentum start from $p_0 = 0$, go through positive values of momentum up to $p_{M/2} = p_{max} - \Delta p$, then discontinuously shift to $p_{M/2+1} = -p_{max}$ and go through all negative values.[62]

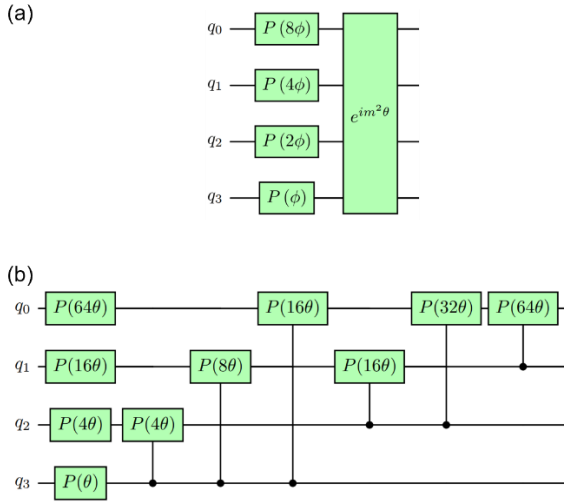

**Figure 7:** a) Quantum circuit to implement the kinetic energy operator in the momentum representation for a 4-qubit system; b) Quantum circuit to implement the quadratic term of the kinetic energy operator in part (a). Black dots indicate control qubits of the two-qubit gates.

### II-F. Initialization of a Compact Wave Packet

Every quantum simulation begins with the initial wave packet, which serves as the starting point for the system's evolution. A simple step-like wave packet, shown in Fig. 8a, is particularly useful due to its straightforward initialization, requiring only a shallow quantum circuit, as illustrated in Fig. 8b for 4 qubits.

At the beginning of any quantum calculation the quantum computer, by default, is set to the initial state $|0000\rangle$, which in our case corresponds to all probability restricted to the first point of the grid: $|\psi_0|^2 = 1$ and $|\psi_m|^2 = 0$ for all $m > 0$. The quantum

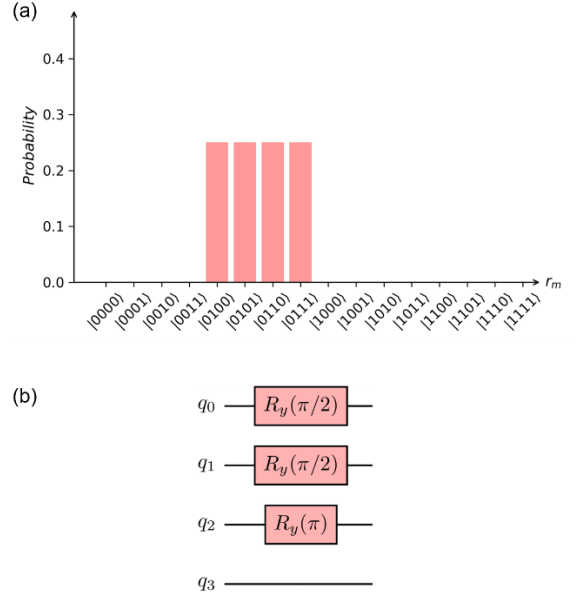

**Figure 8:** a) A step-like initial wave packet in the case of four qubits, suitable for the double-well tunneling problem shown in Fig. 1; b) Quantum circuit to initialize this probability distribution in the 4-qubit case.

circuit given in Fig. 8b is then applied to this initial state as follows:

$$I \otimes R_y(\pi) \otimes R_y\left(\frac{\pi}{2}\right) \otimes R_y\left(\frac{\pi}{2}\right) \begin{pmatrix} 1 \\ 0 \\ 0 \\ 0 \\ 0 \\ 0 \\ 0 \\ 0 \\ 0 \\ 0 \\ 0 \\ 0 \\ 0 \\ 0 \\ 0 \\ 0 \end{pmatrix} = \begin{pmatrix} 0 \\ 0 \\ 0 \\ 0 \\ \frac{1}{2} \\ \frac{1}{2} \\ \frac{1}{2} \\ \frac{1}{2} \\ 0 \\ 0 \\ 0 \\ 0 \\ 0 \\ 0 \\ 0 \\ 0 \end{pmatrix}$$

(23)

This operation effectively spreads out the probability evenly over the second quarter of states in the register. Namely, $|\psi_m|^2 = 1/4$ for $m = 4, 5, 6, 7$, but is zero otherwise, thereby modeling a step-like initial wave packet shown in Fig. 8. This is employed in the next section for the study of quantum tunneling through a barrier. A quantum circuit for a general $n$-qubit encoding of the step-like wave packet is given in the supplementary material. In all these circuits we employ the so-called $Y$-rotation gate given by:

$$R_y(\varphi) = \begin{vmatrix} \cos\left(\frac{\varphi}{2}\right) & -\sin\left(\frac{\varphi}{2}\right) \\ \sin\left(\frac{\varphi}{2}\right) & \cos\left(\frac{\varphi}{2}\right) \end{vmatrix}$$

These gates, unlike the phase gate, affect the probability distribution, moving it between the states of the qubit.

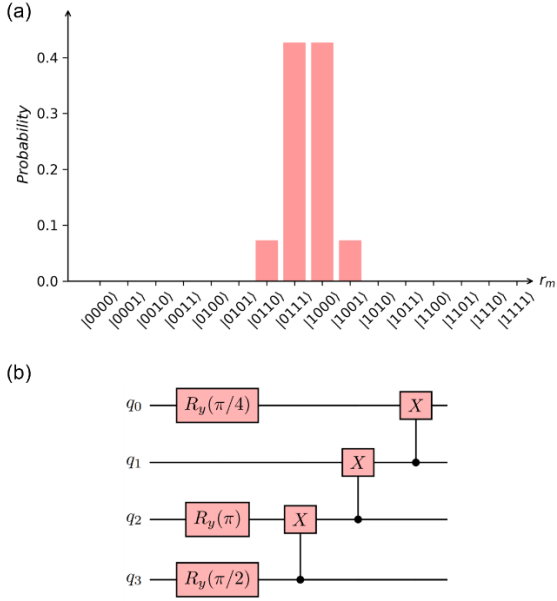

**Figure 9:** a) Gaussian-like initial wave packet in the case of four qubits; b) Quantum circuit to initialize this probability distribution in the 4-qubit case.

A Gaussian wave packet is commonly selected due to its mathematical convenience and physical significance. IBM's Qiskit offers built-in tools, such as the $qc.initialize()$ method, for the initialization of a general wavefunction that can in principle be used to initialize the gaussian wave-packet. However, this approach typically generates a very deep quantum circuit and thus is highly susceptible to hardware noise, resulting in substantial errors. To mitigate this issue, we designed a more efficient, shallower quantum circuit. A 4-qubit implementation of this circuit is shown in Fig. 9b, for the initialization of a compact Gaussian-like wave packet, shown in Fig. 9a. This wave packet is employed in the modeling of the free particle and harmonic oscillator problems. A quantum circuit for a general $n$-qubit encoding of the gaussian-like wave packet is given in the supplementary material. The algorithm requires $n-1$ two-qubit gates for an $n$-qubit system, so, the circuit depth scales linearly with $n$. Alternatively, one can prepare a Gaussian initial state using the variational quantum eigensolver (VQE) approach.[39]

### III. RESULTS AND DISCUSSION

### III-A. Benchmark calculations using classical emulator of quantum computer

First, in order to test that our quantum algorithms and our Qiskit codes are accurate, we conducted a benchmark study by running these codes on a classical emulator of a quantum computer and compared these results with the results of traditional classical algorithms and codes executed on a usual PC. These tests covered three different problems as described next. For each problem we caried out a comprehensive convergence study to ensure that our simulations are converged with respect to the time step $\Delta t$ and the position step-size $\Delta r$. In all cases the range of the coordinate space was $0 \leq r \leq 5$ Bohr. The reduced mass of the OH radical was used, $\mu = 0.9412$ amu, since this was a model system for several quantum computing studies in the past,[63] and the number of propagation time steps was $N = 100$.

*Evolution of a free-particle wave packet:* In this problem we set $V = 0$, so that only a kinetic energy operator is included in the propagator. A grid of 256 points was used and the initial Gaussian wave packet

$$\psi(r) = Ae^{-\left(\frac{r-r_s}{a}\right)^2 + irp_s}$$

with width parameter $a = 0.25$ Bohr was placed at $r_s = 1$ Bohr and was given a momentum $p_s = 30$ a.u. in the positive direction. The propagation time step of $\Delta t = 1.5$ a.u. was used and the total simulation time was $t_{\text{fin}} = 150$ a.u. During the simulation, the wave packet moved through the grid to its final position near $\langle r \rangle = 3.62$ Bohr and had significantly spread, increasing its width, computed as

$$\sigma = \sqrt{\langle r^2 \rangle - \langle r \rangle^2},$$

from $\sigma = 1.13$ Bohr at $t = 0$ to $\sigma = 3.52$ Bohr at the final moment of time. The animation (movie) of this simple process is available from the supplementary material. Importantly, the results of the traditional wave packet propagation method and of the quantum algorithm (executed on a classical simulator of 8 qubits) were in excellent agreement, showing differences on the order of $10^{-13}$ for average position $\langle r \rangle$ and $10^{-12}$ for width $\sigma$, which is within expected numerical error. Also, as one can see from the animation A1, the entire probability distributions $|\psi(r)|^2$ predicted by the two methods were in perfect agreement, during the entire time of the simulation (see Fig. S6a).

*Quantum tunneling through the barrier:* In this case we used a double-well potential (like in Fig. 1) with $V_{min} = -17$ milli Hartree, represented by 128 grid-points or 7 qubits. A uniform initial wave packet was placed into the left well (similar to Fig. 8) at an energy of 0.46 milli Hartree below the barrier top, and was allowed to tunnel into the right well during the total simulation time $t_{fin} = 300$ a.u. The propagation time step of $\Delta t = 3$ a.u. was used. The tunneling probability (into the right well) can be computed as the reduction of probability in the initial (left) well, namely:

$$p = 1 - \sum_{m=M/8}^{5M/8-1} |\psi(r_m)|^2$$

In this computational experiment we found that, during the first 50 a.u. of time the tunneling probability was increasing monotonically up to $p \sim 1\%$, and then kept increasing in a more complex manner reaching $p \sim 3.5\%$ near the end of simulation (see supplementary material). Again, the quantum algorithm run on a simulator was in perfect agreement with traditional propagation method, both in terms of the tunneling probability time dependence $p(t)$ (see Fig. S6b) and the time evolution of the probability amplitude $|\psi(r)|^2$. Readers are encouraged to watch animation A2 of this process available in supplementary material.

We also would like to note that the use of the Fourier transform technique for applying the kinetic energy operator assumes periodic boundary conditions, which means that the tunneling from the initially populated well (the left well in Fig. 1a) happens simultaneously in two directions through two barriers, one to the left and the other to the right of the well. For possible scattering applications a complex absorbing potential needs to be placed at large values of $r$.

*Vibrations of harmonic oscillator:* Here we used a parabolic potential (similar to Fig. 4) centered at $r_{eq} = 2.5$ Bohr and represented by 256 grid-points or 8 qubits, with harmonic frequency $\omega = 3978.6$ cm$^{-1}$ or approximately 18 milli Hartree. The initial Gaussian wave packet with width parameter $a = 0.36$ Bohr, which is about 40% wider than the ground state solution in this potential, was displaced off the equilibrium position to $r_s = 1.5$ Bohr, which corresponds to the initial total energy of approximately $32\omega$. The propagation time step of $\Delta t = 11$ a.u. was used and the total simulation time was $t_{fin} = 1100$ a.u. In the animation of this simulation (see the movie A3 in supplementary material) one can observe that this wave packet evolves through three periods of molecular vibrations (bond stretching and compression): changing its shape periodically, spreading at the turning points and refocusing at the equilibrium position. Importantly, the results of the quantum algorithm are in excellent agreement with that based on the traditional simulation method (see Fig. S6c). Deviations of average bond length, computed as

$$\langle r \rangle = \sum_{m=1}^{M} r_m |\psi(r_m)|^2$$

were on the order of $10^{-13}$, which is within numerical accuracy.

### III-B. Testing the actual present day quantum hardware available to the community

Since the benchmark calculations with quantum algorithms on a classical emulator of the quantum computer were successful, we tried to move further and tried to run the same algorithms on the actual quantum hardware. The same three test problems were simulated, but several input parameters were modified to simplify the task. Namely, the number of qubits was reduced to 5, and the corresponding number of grid points to 32. The number of time-steps was reduced to $N = 8$. Several other input parameters were modified as explained below. The results obtained on a classical emulator of quantum hardware were used as a benchmark. Those are presented by green symbols and lines in Fig. 10.

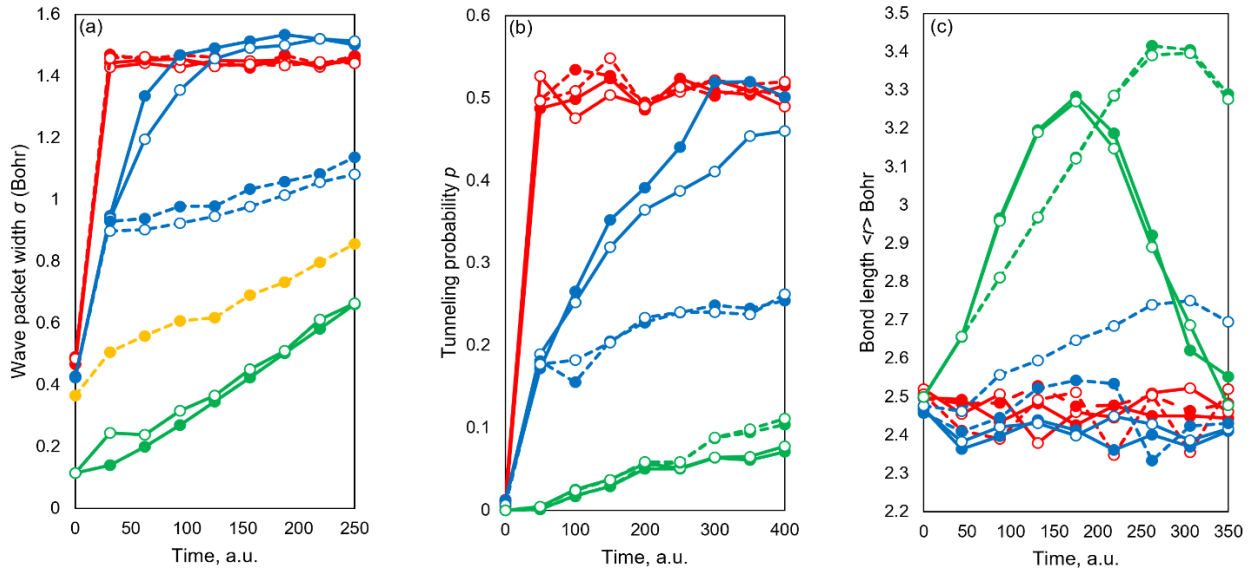

**Figure 10:** Three panels illustrate results from three test problems described in the text: (a) Evolution of a free-particle wave packet, (b) Quantum tunneling through a barrier, and (c) Vibrations of a harmonic oscillator. Green symbols represent results from the classical emulator of the quantum computer, blue symbols correspond to IBM Torino (newer *Heron* processor), and red symbols to IBM Brisbane (older *Eagle* processor). Yellow symbols in panel (a) correspond to IonQ *Aria* 1 quantum computer. Filled symbols denote results obtained using exact QFT, while empty symbols indicate the results of approximate QFT. Solid lines represent the multiple-step approach, whereas dashed lines correspond to the single-step approach.

*Evolution of a free-particle wave packet:* A compact Gaussian-like wave packet as in Fig. 9 was placed in the middle of the $r$-domain, without any initial momentum. We used $\Delta t = 31.25$ a.u. and the time of simulation was $t_{\text{fin}} = 250$ a.u., during which the average width of the wave packet in the benchmark simulations increased from $\sigma = 0.12$ to $0.66$ Bohr. Figure 10a summarizes the results of these tests.

*Quantum tunneling through the barrier:* A well depth of $V_{min} = -5$ milli Hartree was used, and the energy of the initial wave packet was 0.89 milli Hartree below the barrier top. The time step was $\Delta t = 50$ a.u. and the total simulation time was $t_{\text{fin}} = 400$ a.u., during which the tunneling probability in the benchmark simulations reached $p \sim 7\%$, as shown in Fig. 10b. The final snapshot of the wave packet evolution for this problem one can see in the graphical abstract.

*Vibrations of harmonic oscillator:* Here we used a parabolic potential centered at $r_{eq} = 3$ Bohr. A Gaussian-like initial wave packet was displaced off the equilibrium position to $r_s = 2.5$ Bohr, which corresponds to the initial total energy of approximately $4\omega$. The time step was $\Delta t = 43.75$ a.u. and the total simulation time was $t_{\text{fin}} = 350$ a.u. which approximately corresponds to one period of vibration on a classical emulator, as one can see from Fig. 10c.

First, let's discuss the quality of the initial state preparation on the actual quantum hardware, reflected in Fig. 10 by the first point at $t = 0$, which represents the read-out of the final state of the quantum computer right after the initial state was prepared (*i.e.*, without any time-propagation steps).

From Fig. 10a one can clearly see that the values of wave packet width $\sigma$ set up on the quantum hardware (blue and red symbols at $t = 0$) significantly deviate from the benchmark calculations on a classical emulator (green symbols at $t = 0$). These errors will propagate in time and will contribute to the overall error of the result at the final moment of time $t_{\text{fin}}$. From Fig. 10a one can also see that the value of error depends on the type of quantum hardware. For example, we found that the initial state preparation errors are visibly smaller when using the IonQ *Aria* 1 quantum computer (yellow symbols), compared to IBM *Torino* (blue symbols) and IBM *Brisbane* (red symbols). We also tested the *Ankaa*-3 machine provided by Rigetti Computing and found its results to be comparable to those obtained on the *Eagle* processor of IBM. Therefore, they were not included in our analysis.

Now, let's assess the quality of solution after the first time-propagation step, at $t = \Delta t$. From Fig. 10 one can see that when the older *Eagle* processor hardware is used, such as IBM *Brisbane*, then after only one propagation step the results become corrupted (red symbols and lines). The resulting values of $\sigma$, $\langle r \rangle$ and $p$ correspond to a uniform population of all qubit states, which in turn represents a flat probability distribution over the entire $r$-grid of 32 points. When the newer *Heron* processor is used, such as IBM *Torino*, the results are considerably better (blue symbols and lines), although they still deviate from the benchmark data (green symbols and lines). The results obtained from IonQ *Aria* 1 (yellow symbols and lines) are even more accurate, showing only half the deviation from the benchmark data compared to IBM *Torino*. At this point one can draw a conclusion that the older *Eagle* processors are too noisy for a practical implementation of the kinetic energy operator (using QFT) and/or the potential energy operator of the harmonic oscillator. It appears that these quantum circuits are too deep for these older devices. The results obtained on the *Ankaa*-3 machine by Rigetti were, again, comparable to those obtained on the IBM *Eagle* processor.

For this reason, we decided to try a simplified version of QFT available in Qiskit. In this approximation the QFT circuit is made shallower by neglecting less significant controlled-phase shifts, such as $CP(\pi/8)$ in the 4-qubit example of Fig. 6. In Figure 10 the results of exact QFT are given by filled symbols, while the results of the approximate QFT are given by empty symbols (of all colors). One can see that on the older quantum hardware (red symbols) this approximation did not make any difference for all three test problems considered here. However, on the IBM *Torino* a visible improvement was observed when the approximate QFT was used. Namely, in Fig. 10a (both solid and dashed lines), in Fig. 10b (focus on solid lines) and in Fig. 10c (focus on dashed lines) the empty blue symbols are closer to empty green symbols, whereas the filled blue symbols are further from filled green symbols. This observation is encouraging. In what follows we will focus on the results obtained using the newer quantum hardware, such as IBM *Torino*.

It should be pointed out that on the actual quantum hardware only one (final) read-out is possible, in contrast to the classical computers or classical emulators of quantum hardware, where one can proceed with calculations to the next step, after the result of the previous step is recorded. Therefore, in order to obtain the results of time propagation as a function of time, as in Fig. 10, we had to run eight independent calculations on the actual quantum hardware: one time-step followed by read out, then independently two time-steps followed by read out, and so on, up to eight time-steps followed by read out.

From the time-dependencies presented in Figs. 10a and 10b, it follows that although the results obtained on the IBM *Torino* resemble the benchmark results during the first few time-steps, they quickly deteriorate (solid blue lines) and, after eight time-steps, they eventually approach the uniform distribution observed on the IBM *Brisbane* (red lines). Searching for possible ways of improvement, we tried an alternative method of reconstructing these time dependencies. Instead of making several time-steps of the same length $\Delta t$, we tried to run several single-step calculations. Namely, the second point on the graph was obtained from the calculation with a single time-step $2\Delta t$, the third point on the graph with $3\Delta t$ and so on, up to the last point that was obtained from a single-step calculation with time step $8\Delta t$. This approach sacrifices numerical accuracy of the propagator (because the time-steps are larger) but makes the quantum circuits shorter by minimizing the overall number of gates. The results of this approach are presented in Fig. 10 by the dashed lines.

Comparing dashed *vs* solid lines in Fig. 10 one concludes that, on the actual quantum hardware we tested, the single-step approach is superior to the multiple step approach, for the reason stated above. In particular, the harmonic oscillator problem, Fig. 10c, appears to be the hardest due to the complexity of potential energy operator. In this case, some qualitative agreement between the results obtained on quantum hardware and those of the benchmark calculations, was obtained only if both a simplified QFT and a single-step propagation approach are employed (dashed lines, empty symbols, blue vs green). Although the allocation we had on IonQ was sufficient to run only one simple test problem (the evolution of a free wave packet, yellow symbols in Fig. 10a), we found that this quantum processor is visibly better than all other processors tested in this work. This is consistent with the higher accuracy of two-qubit gates enabled by the trapped ion technology that possesses all-to-all connectivity.[64–66]

The supplementary material contains animations of the wave packet evolution obtained on the quantum hardware (IBM *Torino*), for three test problems considered here, using the best quantum algorithm we found for each case, which is the single-step propagation method combined with approximate QFT. Readers are encouraged to watch these movies.

## IV. CONCLUSIONS

In this paper we outlined all components of a quantum algorithm needed for the practical realizations of quantum molecular dynamics simulations on a quantum computer. A grid representation of the wavefunction was employed, mapped onto the qubits of the quantum computer. A split-operator method was implemented using the QFT for the kinetic energy operator. Potential energy operators were implemented for a double-well and a harmonic oscillator potential. We found that Qiskit is a platform quite convenient for these simulations, well-developed and periodically updated. Our quantum codes were rigorously tested by running them on the emulator of quantum hardware and comparing their results against results obtained by traditional "classical" methods. Potential users can find this code in the supplementary material for this paper.

The success of running this code on the actual quantum hardware depends on the specific hardware used. Some of the newer quantum hardware available today, such as IonQ *Aria* and IBM *Torino*, permits us to obtain results that are semi-quantitatively similar to the results of the benchmark calculations, and this is encouraging. However, these results remain too noisy for practical implementation of deep quantum circuits, such as those for the kinetic energy operator (using QFT) and/or the potential energy operator for the harmonic oscillator. To improve performance, the quantum circuits need to be simplified, but without significant loss of accuracy, as we have not yet reached the era of fault-tolerant quantum computers. Therefore, the development of new quantum algorithms that could apply the kinetic energy operator with shallower circuits [67–69] is of primary importance for the progress of this field.

## SUPPLEMENTARY MATERIAL

The supplementary material contains one pdf file with diagrams of the quantum circuits used in this work but for an arbitrary number of qubits $n$ and a Qiskit program to implement the benchmark tests on a classical emulator of quantum hardware as described in Section III-A of the paper. It also contains six animations (movies) to visualize the dynamics of wave packets in the three test problems we considered here. Animations A1, A2 and A3 correspond to the benchmark tests on a classical emulator of quantum hardware as described in Section III-A of the paper. Animations A4, A5 and A6 correspond to the calculations run on the actual quantum hardware (IBM Torino) as described in Section III-B of the paper.

## ACKNOWLEDGMENTS

This research was supported by NSF ExpandQISE program, grant number OSI/OMA - 2328489. DB acknowledges the support of Habermann-Pfletschinger Research Fund. TK acknowledges the support of Quantum Computing Summer School at LANL. BKK acknowledges that part of this work was done under the auspices of the U.S. Department of Energy under Project No. 20240256ER of the Laboratory Directed Research and Development Program at Los Alamos National Laboratory. Los Alamos National Laboratory is operated by Triad National Security, LLC, for the National Nuclear Security Administration of the U.S. Department of Energy (Contract No. 89233218CNA000001).

## CONFLICTS OF INTEREST

The authors have no conflicts to disclose.